**Space charge limited conduction with exponential trap distribution in reduced graphene oxide sheets**

Daeha Joung[1,2], A. Chunder[1,3], Lei Zhai[1,3], and Saiful I. Khondaker[1,2] *

[1] Nanoscience Technology Center, [2] Department of Physics, [3] Department of Chemistry, University of Central Florida, Orlando, Florida 32826, USA.

* To whom correspondence should be addressed. E-mail: saiful@mail.ucf.edu

**ABSTRACT**

We elucidate on the low mobility and charge traps of the chemically reduced graphene oxide (RGO) sheets by measuring and analyzing temperature dependent current-voltage characteristics. The RGO sheets were assembled between source and drain electrodes via dielectrophoresis. At low bias voltage the conduction is Ohmic while at high bias voltage and low temperatures the conduction becomes space charge limited with an exponential distribution of traps. We estimate an average trap density of $1.75 \times 10^{16}$ cm$^{-3}$. Quantitative information about charge traps will help develop optimization strategies of passivating defects in order to fabricate high quality solution processed graphene devices.

Nanostructures of graphene oxide (GO) have created a lot of attention as it provides a pathway to produce large quantities of graphene sheets in solution at low cost.[1-4] The easy processibility of GO and compatibility with various substrates including plastics makes them an attractive candidate for high yield manufacturing of graphene based electronic devices. However, GO is an electrically insulating material consisting of a large number of C-O bonds. Removal of C-O bonds by chemical and/or thermal reduction technique produces reduced graphene oxide (RGO) sheets and allows one to restore the electrical properties. RGO sheets have been used for field effect transistors,[1-4] chemical sensors,[5] organic solar cells,[6] as well as transparent electrodes in photovoltaic devices.[7-8] However, the electrical conductivity and field effect mobility values for RGO sheets are much inferior to that of pristine graphene. This has been attributed to a large amount of disorder present in the RGO sheets. Disorder in RGO sheets have mostly been characterized with temperature dependence of resistivity where variable range hopping and activated hopping have been observed.[2,9] Such a study does not provide detailed information about the nature and density of charge traps in the system. A detailed knowledge of charge conduction and quantitative information about charge traps is important for further optimization of reduction techniques and synthetic strategies for increasing conductivity and mobility of RGO towards pristine graphene. Similar strategies have been successfully used for passivating defects in silicon to fabricate high quality complementary metal oxide semiconductor circuits.

In this paper we present measurements and analysis of temperature dependent current density ($J$)-voltage ($V$) characteristics to elucidate on the nature and density of charge traps in the RGO sheets. The GO sheets suspended in water were reduced chemically in the solution and then assembled between prefabricated gold source and drain electrodes via ac dielectrophoresis (DEP). We show that the $J$-$V$ characteristic of the RGO devices measured at different temperatures (295 to 77 K) follow power law behavior, $J \propto V^m$. At low bias, the conduction is Ohmic with *m*=1, while at higher bias voltages *m* increases from 2 to 3 with reducing temperature signifying space charge limited conduction (SCLC) with a transition from trap free (TF-SCLC) regime at room temperature to exponentially distributed trap (EDT-SCLC) regime at low temperatures. We estimate an average trap density of $1.75 \times 10^{16}$ cm$^{-3}$ for our samples.

RGO sheets were synthesized through a chemical reduction of GO in solution prepared by modified Hummers method.[10] Oxidized graphite in water was ultrasonicated to achieve GO monolayer sheets followed by centrifugation for 30 minutes at 3000 rpm to remove any unexfoliated oxidized graphite. The pH of GO dispersion in water (0.1 mg/mL) was adjusted to 11 using a 5% ammonia aqueous solution. 15 μL of hydrazine solution (35% in DMF) was then added to the mixture. The mixture was heated at 95-100 $^0$C for one hour and cooled to room temperature. Devices were fabricated on heavily doped silicon (Si) substrates capped with a thermally grown 250 nm thick SiO$_2$ layer. Source and drain electrode patterns with a channel length of 500 nm and channel width of 500 nm were defined by means of electron beam lithography (EBL) and thermal deposition of 5 nm thick Cr and 20 nm thick Au followed by lift off.

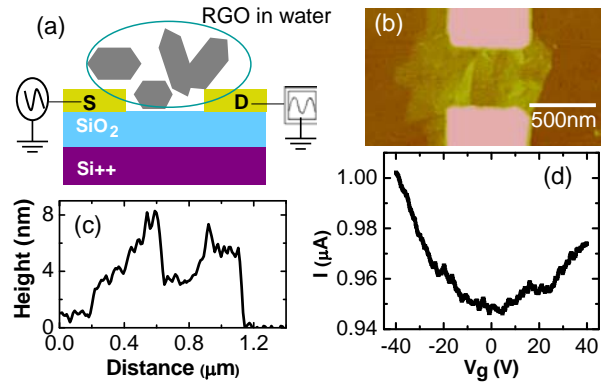

The RGO sheets were assembled between the source and drain electrodes via ac dielectrophoresis (DEP) in a probe station under ambient condition. Figure 1 (a) shows a cartoon of the DEP set up. Details of the assembly were presented in our previous publication.[11] In brief, a small drop of RGO (2 μL) solution was placed onto source and drain electrodes pattern. An *AC* voltage of approximately 3 $V_{P-P}$ at 1 *MHz* was applied with a function generator for 20-30 seconds. The *AC* voltage gave rise to a time averaged DEP force which causes the RGO sheets to align between the source and drain electrodes. After the assembly, the solution droplet was blown off by nitrogen gas. Figure 1(b) shows tapping-mode atomic force microscope (AFM) image of a representative devices with the height analysis shown in Fig. 1(c). It can be seen that the thickness varies from 2 nm to 8 nm in the channel indicating that up to 8 layers of RGO sheets have been assembled in the channel. The thickness is lower at the edges, demonstrating one or two layers of graphene sheet near the edge while the thickness is higher in the middle of the channel due to overlap of several individual sheets or folding of sheets. The devices were then bonded and loaded into a variable temperature cryostat. Temperature dependent electronic transport measurements from 295 K to 77 K were performed using a Keithley 2400 source-meter, and a current preamplifier (DL 1211) capable of measuring sub-pA signal interfaced with LABVIEW program. A total of nine devices were investigated.

FIG. 1. (Color online) (a) Cartoon of DEP assembly set-up. (b) Tapping-mode AFM of a RGO device assembled via DEP. Scale bar represents 500 nm. (c) Height profile of the AFM image shown in Fig. 1 (b). (d) Room temperature transfer characteristic of device A.

Figure 1 (d) shows the room temperature transfer characteristics of a representative RGO device (device A) where the current (*I*) is plotted as a function of gate voltage ($V_g$) measured in vacuum. The gate voltage was scanned from -40 to +40 V with a fixed source-drain voltage of V=0.5 V. The device shows an ambipolar field effect transistor (FET) behavior while. The room temperature hole mobilites were calculated as 0.71 cm$^2$/Vs.

Figure 2 (a) and (b) show the *I-V* characteristics at various temperatures measured from -5 V to 5 V with $V_g= 0$ for device A. The magnitude of the applied bias was limited to 5V to avoid electrical breakdown of the RGO devices. The *I-V* curves are highly symmetric at all temperatures. In addition, the *I-V* curve became increasingly nonlinear with decreasing temperature. The nonlinearity in *I-V* curves in disordered semiconductor systems has been explained using (i) Schottky barrier (SB) between metal electrode and semiconductor, (ii) Fowler-Nordheim (FN) tunneling, and (iii) space charge limited conduction (SCLC). In general, the *I-V* curve of a metal-semiconductor-metal junction should be highly asymmetric if the Schottky barrier is dominated. However, our *I-V* curves are highly symmetric (see Fig. 2 (b)) giving evidence that charge transport is not dominated by SB. Additionally, the attempt to fit our data with the SB was not successful (not shown here). Fowler-Nordheim tunneling can be ruled out as in FN model,[12] the *I-V* curves are independent of temperature while our *I-V* curves are highly sensitive to temperature.

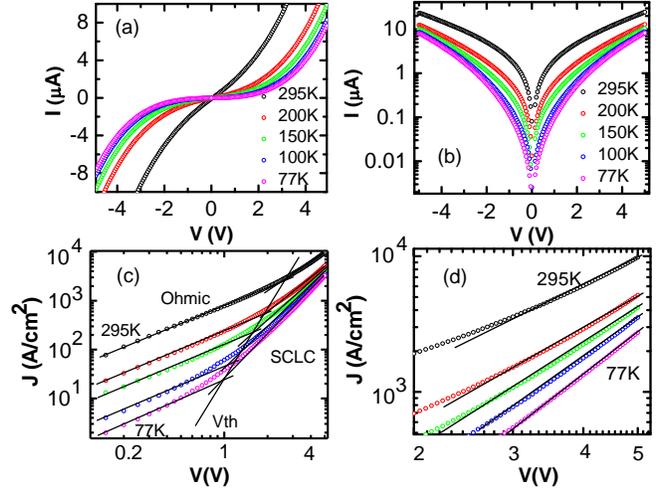

FIG. 2. (Color online) (a) Current – Voltage *(I-V* characteristics at different temperatures for device A. (b) *I-V* curves at different temperatures plotted in semi log scale to clearly show highly symmetric nature of the curves. (c) Current density *(J)* versus *V* plotted in a log-log scale for different temperatures. The symbols are the experimental points while the solid lines are fits to $J \propto V^m$. Two regions are separated by $V_{th}$. For $V < V_{th}$, $m=1$ and the conduction is Ohmic. $V > V_{th}$ the conduction is due to SCLC with exponentially distributed trap states. (d) Expanded view of the SCLC regime. For clarity, *J* at 77K was divided by 1.2.

We now discuss SCLC mechanism. SCLC occurs in low mobility semiconductors when injected charge density exceeds the intrinsic free carrier density of the material. Analysis of the *J-V* characteristics using the SCLC model is one of the experimental methods for the detection of charge trap states in disordered semiconductors. In the absence of any trap states or when trap states do not dominate the transport, the *J-V* characteristics are described using, $J = 9\varepsilon_0\varepsilon_r\mu V^2/8d^3$ where $\varepsilon_0$ is the permittivity of free space, $\varepsilon_r$ the dielectric constant of the RGO, $\mu$ is the charge carrier mobility, *d* is the spacing between electrodes.[13] However, in presence of trap states that are exponentially distributed in energy, the *J-V* relationship is given by $J = (\mu N_v / q^{l-1})((2l+1)/(l+1))^{l+1}((\varepsilon\varepsilon_0/N_t)(l/(l+1)))^l (V^{l+1}/d^{2l+1})$ where, *l* is an exponent and should be greater than 1, *q* is the electronic charge, $N_v$ is effective density of states, $N_t$ is the trap density.[14] By plotting *J* and *V* in log-log scale one can determine the value of exponent m. For trap free (TF-SCLC) regime, *m* =2 while for exponentially distributed trap (EDT-SCLC) regime *m=l+1>2*.

In order to examine whether our data can be explained using SCLC model, we have plotted, in Fig. 2 (c), *J* versus *V* curves at different temperature in a log-log scale for sample A. The dotted symbols are the experimental data points and the solid lines are a fit to $J \propto V^m$. At low voltages, *m* equals to 1 at all temperatures signifying Ohmic conduction. However, at higher voltages, *m* deviates from 1. $V_{th}$, the onset voltage where log *J* – log *V* curves begin to inflect to higher slope region, shifts to lower voltage with decreasing temperature. For $V>V_{th}$, *m*=2 at room temperature implying TF-SCLC regime. However, as the temperature is reduced, the value of m for $V>V_{th}$ is increased to 2.3 at 200 K, 2.6 at 150 K, 2.8 at 100 K and 3 at 77 K. The *J-V* curves for higher values of *V* are more clearly shown in Fig. 2 (d). For these temperatures, the conduction is governed by traps that are exponentially distributed in energy (EDT-SCLC).

A transition from TF-SCLC regime at room temperature to EDT-SCLC regime at low temperature can occur for the following reason: At room temperature, the free career density is higher than the trap density. However, as the temperature is reduced, free career density is also reduced and the trap states started to dominate by localizing charge carriers and we enter EDT-SCLC regime with *m*>2.

More quantitative information about traps can be obtained by extrapolating *J-V* curves in voltage. If charge traps are distributed in energy, they will be gradually filled with increasing electric field at all temperatures and at a certain critical voltage, all traps will be filled. According to Kumar et al [15] this critical voltage is independent of temperature and is given by $V_c = qN_t d^2/2\varepsilon_r \varepsilon_0$. By extrapolating the *J-V* curves we can obtain values for $V_c$ and calculate trap density $N_t$. In Fig. 3, we extrapolate the log *J* and log *V* characteristic at higher bias voltages for all temperatures except room temperature. The $V_c$=10 V was found for device A. In order to calculate the trap density from $V_c$, we need to know the dielectric constant ($\varepsilon_r$) of chemically reduced graphene oxide. The value of $\varepsilon_r$ can be calculated by using; $\varepsilon_r = n^2 - k^2$, where *n* and *k* are refractive index and extinction coefficient of RGO film respectively. Using the values of *n* and *k* in the thermally reduced GO,[16] we calculate $\varepsilon_r$ = 3.5. Using this value, we obtain trap density $N_t$ = 1.54x10$^{16}$ cm$^{-3}$ for A.

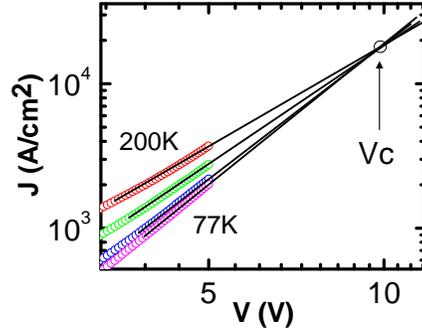

FIG. 3. (Color online) Extrapolation of the *J-V* curves at low temperatures plotted in log-log scale for device A. From this plot, we determine the temperature independent crossover voltage $V_c$=10 V.

Similar EDT-SCLC regime at low temperatures was observed in all nine samples that we have measured. In Table I, we summarize all measured RGO devices where we tabulate room temperature hole mobility, $V_c$ and $N_t$ values. All measured devices show similar trap density. The average trap density was about 1.75 x 10$^{16}$ cm$^{-3}$. The value of trap density is similar to what has been reported for CdSe nanocrystal and amorphous polymer semiconductors.[17,18]

An exponential distribution of trap in energies is expected for the traps originating from surface defects and structural disorders.[19] It has been shown that GO consists of a lot of C-O bonds and that complete removal of C-O bonds via chemical and or thermal reduction is not possible.[20] Removal of C-O bonds increases $sp^2$ concentration of C-C bond increasing the conductivity. However the remaining oxygen cause a large number of $sp^3$ bonding creating disordered regions with trap states. Additionally GO also undergoes structural changes due to loss of oxygen during reduction creating topological defects. In addition, line defects such as wrinkles and folding of GO sheet also create defects. All these defects make RGO a rough and defective graphene sheet. Because of the presence of the large number of trap states caused by these defects, charge in the gate would induce charge in the traps rather than free charge carrier in the semiconducting RGO sheets causing a low mobility in these sheets. Recent studies of structural properties of GO show how the *sp$^2$* fraction of C-C bond changes with different stage of reduction.[20] It will be interesting to study the density of defects/traps with such controlled reduction strategies using technique presented in this paper.

Table I. The values for room temperature hole mobility, cross over voltage ($V_c$), and trap densities ($N_t$) for all nine samples that we measured.

| Sample | Hole mobility (cm²/V) | $V_c$ (V) | $N_t$ (cm$^{-3}$) |
|---|---|---|---|
| A | 0.71 | 10.0 | 1.54 X 10$^{16}$ |
| B | 0.67 | 10.0 | 1.54 X 10$^{16}$ |
| C | 0.72 | 12.5 | 1.90 X 10$^{16}$ |
| D | 0.48 | 10.0 | 1.54 X 10$^{16}$ |
| E | 0.38 | 11.5 | 1.80 X 10$^{16}$ |
| F | 0.96 | 12.0 | 1.85 X 10$^{16}$ |
| G | 0.71 | 11.5 | 1.80 X 10$^{16}$ |
| H | 1.05 | 12.5 | 1.93 X 10$^{16}$ |
| I | 0.68 | 11.0 | 1.70 X 10$^{16}$ |

In conclusion, we elucidated on the low mobility and density of charge traps in RGO sheets by measuring and analyzing temperature dependent current-voltage characteristics. The GO sheets suspended in water were reduced chemically in the solution and then assembled between prefabricated gold source and drain electrodes via ac dielectrophoresis (DEP). We show that at low bias voltage the conduction is Ohmic while at high bias voltage the conduction becomes SCLC. At room temperature, the conduction is governed by trap free SCLC while at a lower temperature it is dominated by traps that are exponentially distributed in energy. We estimated an average trap density of $1.75 \times 10^{16}$ cm$^{-3}$. Quantitative studies of charge traps using SCLC presented in this work will facilitate further development of strategies for the chemical modification of RGO surfaces to passivate traps and thereby improve the carrier transport by providing quantifying evidences of the developed strategies of passivating defects in order to fabricate high quality solution processed graphene devices.

**Acknowledgement:**
This work has been partially supported by US NSF under grants ECCS 0748091 to SIK and DMR 0746499 to LZ.